\documentclass[aoas,MSNbibl,nameyear,dvips]{arximspdf}
\usepackage{dcolumn}
\usepackage{graphicx}

\doi{10.1214/12-AOAS535} 
\volume{6}
\issue{3}
\pubyear{2012}
\firstpage{1118}
\lastpage{1133}

\makeatletter
\newcolumntype{d}[1]{D{.}{.}{#1}}
\newcommand{\eqref}[1]{(\ref{#1})}

\newcommand{\I}{\mathbf{I}}
\newcommand{\J}{\mathbf{J}}

\newcommand{\bt}{\bolds{\theta}}
\makeatother

\begin{document}
\begin{frontmatter}

\title{The importance of distinct modeling strategies for gene and
gene-specific treatment effects in hierarchical models for microarray data\thanksref{T1}}
\runtitle{Modeling gene and gene-specific treatment effects}

\thankstext{T1}{Supported by the National Research Initiative of the
USDA-CSREES Grant 2008-35600-18786.}

\begin{aug}
\author[A]{\fnms{Steven P.} \snm{Lund}\ead[label=e1]{lunds@iastate.edu}}
\and
\author[B]{\fnms{Dan} \snm{Nettleton}\corref{}\ead[label=e2]{dnett@iastate.edu}}
\runauthor{S. P. Lund and D. Nettleton}
\affiliation{Iowa State University}
\address[A]{Department of Statistics\\
Iowa State University\\
3418 Snedecor Hall\\
Ames, Iowa 50011-1210\\
USA\\
\printead{e1}} 
\address[B]{Department of Statistics\\
Iowa State University\\
2115 Snedecor Hall\\
Ames, Iowa 50011-1210\\
USA\\
\printead{e2}}
\end{aug}

\received{\smonth{7} \syear{2011}}
\revised{\smonth{1} \syear{2012}}

%
\begin{abstract}
When analyzing microarray data, hierarchical models are often used to
share information across genes when estimating means and variances or
identifying differential expression. Many methods utilize some form of
the two-level hierarchical model structure suggested by
Kendziorski et~al. [\textit{Stat. Med.} (2003) \textbf{22}
3899--3914] in which the first level describes the distribution of
latent mean expression levels among genes and among differentially
expressed treatments within a~gene. The second level describes the
conditional distribution, given a latent mean, of repeated observations
for a single gene and treatment. Many of these models, including those
used in Kendziorski's et~al. [\textit{Stat. Med.} (2003) \textbf{22}
3899--3914] EBarrays package, assume that expression
level changes due to treatment effects have the same distribution as
expression level changes from gene to gene. We present empirical
evidence that this assumption is often inadequate and propose
three-level hierarchical models as extensions to the two-level
log-normal based EBarrays models to address this inadequacy. We
demonstrate that use of our three-level models dramatically changes
analysis results for a variety of microarray data sets and verify the
validity and improved performance of our suggested method in a series
of simulation studies. We also illustrate the importance of accounting
for the uncertainty of gene-specific error variance estimates when
using hierarchical models to identify differentially expressed genes.
\end{abstract}

%

\end{frontmatter}

\section{Introduction}
\label{s:intro}

There are many analytic methods for microarray data that utilize a
hierarchical model to share information across genes when estimating
mean expression levels. A large subset of these methods model
differences in expression levels from gene to gene and differences in
expression levels caused by treatment effects with a single
distribution. Canonical examples of such methods are implemented in the
EBarrays package for R developed by Kendziorski et~al. (\citeyear{Kendziorski}). This work has
been influential as indicated by a variety of recent methods that cite
Kendziorski et~al. (\citeyear{Kendziorski}) and follow their modeling strategy. Examples
include \citet{Newton2004}, \citeauthor{YuanandKendziorski} (\citeyear{YuanandKendziorski,Yuan2006b}),  \citet{Yuan2006a},
\citet{Lo}, \citet{Keles}, \citeauthor{Wei2007} (\citeyear{Wei2007,Wei2008}), \citet{Wu}, \citet{Jensen}, and \citet{Rossell}.

The analytic methods provided in EBarrays are based on two-level
hierarchical parametric models that can be used to analyze data from
experiments with more than two treatment groups and produce posterior
expression pattern probabilities, which can be used to assess the
significance of and classify differential expression of genes. The
first level of the hierarchical model describes the distribution of
latent mean expression levels among genes and among differentially
expressed (DE) treatments within a gene. The second level describes the
conditional distribution, given a latent mean, of repeated observations
for a single gene and treatment.

A necessary user input to models like those included in EBarrays is
a~list of possible expression patterns. In a two-treatment experiment,
the only two expression patterns are equivalent expression and
differential expression. In general, each pattern describes how to
partition the experimental units into groups based on the experimental
conditions or treatments associated with the experimental units. An
analysis based on these models can yield a~gene-specific posterior
probability estimate for each pattern.

The application of hierarchical models to microarray data has many
benefits: ``sharing'' information across genes compensates for having
few replicates, users may define expression patterns of interest
involving two or more experimental conditions, posterior probabilities
assigned to expression patterns are easy to interpret and allow for
easy classification or ranking, and simultaneous analysis of all genes
in a data set greatly reduces the dimensionality of the inference
problem. While the work of Kendziorski et~al. (\citeyear{Kendziorski}) lays a foundation for a
powerful method of microarray analysis upon which many methods have
been developed, there is room to relax assumptions and to improve the
models described.

The main point of this paper concerns the assumption---implied by the
modeling strategy of Kendziorski et~al. (\citeyear{Kendziorski})---that expression changes
across genes have the same distribution as expression changes caused by
treatment effects. This assumption is convenient for computational
reasons but has undesirable consequences. In particular, if expression
differences from gene to gene tend to be larger than treatment effects,
the power to identify differentially expressed genes will be reduced.
Based on our experience with microarray data, we see no reason to
believe that expression differences across genes have the same
distribution as expression differences caused by treatment effects in\vadjust{\goodbreak}
all experiments. Thus, we propose to relax this assumption by adding an
additional level to the hierarchy of Kendziorski's et~al. (\citeyear{Kendziorski})
lognormal--normal (LNN) model. This creates a three-level hierarchical
model that we will call the lognormal--normal--normal (LN$^3$) model.

A secondary point of this paper concerns the assumption of a constant
coefficient of variation used in Kendziorski's et~al. (\citeyear{Kendziorski}) gamma--gamma
(GG) and lognormal--normal (LNN) models, which, for the latter model,
implies an error variance of log expression values that is common to
all genes. This assumption is now widely regarded as untenable. To
address this, \citet{Lo} introduced a method to relax the assumption
of the GG and LNN models, and many methods to estimate gene-specific
error variances for microarray data have been developed.
[See, e.g., \citet{Baldi}, \citet{Lonnstedt}, \citet{Wright}, \citet{Cui}.]
Kendziorski's et~al. (\citeyear{Kendziorski})
EBarrays package includes the LNN-moderated variance
(LNNMV) method, which uses shrunken point estimates of gene-specific
error variances similar to those described by \citet{Smyth}. We
briefly demonstrate that using point estimates of gene-specific error
variances without accounting for their uncertainty produces liberal
posterior pattern probability estimates, which causes underestimation
of the proportion of false positives on a list of significant genes. We
propose a~simple adaptation to the LNNMV method to account for the
uncertainty in gene-specific variance estimates and demonstrate this
corrects the liberal bias in the estimated expression pattern posterior
probabilities. Finally, we combine our proposed three-level
hierarchical modeling strategy with gene-specific error variance
modeling to obtain a more general model denoted LN$^3$MV.

We formally describe the four lognormal based models (LNN, LNNMV,
LN$^3$, and LN$^3$MV) and corresponding analytic methods in Section~\ref{s:Model Descriptions}.
In Section~\ref{s:Evidence Supporting Need for Three-level Hierarchical Models}
we present empirical evidence from two example microarray
data sets that clearly supports our proposed three-level hierarchical
modeling strategy. In Section~\ref{s:Data Analysis} we demonstrate the practical impact of
our suggested adaptations by analyzing data from the two microarray
experiments with several methods. Section~\ref{s:Simulation Study} describes a variety of
simulation studies used to verify the validity and improved power of
our suggested methods. For both real and simulated data sets, the use
of our proposed three-level hierarchal model dramatically increases
power to detect DE genes.

\section{Model descriptions}
\label{s:Model Descriptions}

Throughout this paper, we will use the term ``group'' to denote a set of
equivalently expressed (EE) observations from a single gene. Consider a
microarray data set with expression values for~$J$ genes from each of
$I$ experimental units divided among 2 experimental conditions. If for
gene $j$ there is no difference between the expression distributions
for experimental units under conditions~1 and 2, then the entire set of
$I$ observations forms a single group. If for gene $j$ there is a
difference between the expression distributions for experimental units
under conditions 1 and~2, then the set\vadjust{\goodbreak} of observations from
experimental units under condition 1 forms one group and the set of
observations from experimental units under condition 2 forms a second
group. In general, there is at least one group for every gene and at
most one group for every combination of gene and experimental condition.

Throughout this section, we will use $G_{p}(i)$ to denote the group
(subset of EE observations) to which the $i$th experimental unit
belongs under the $p$th expression pattern. For example, suppose there
is an experiment with~6 experimental units distributed across 3
treatment groups labeled control,~A, and~B. If a researcher aims to
compare each of treatments A and B to the control, then expression
patterns of interest for each gene are $p=1$: control${}={}$A, control${}={}$B;
$p=2$: control${}\not={}$A, control${}={}$B; $p=3$: control${}={}$A, control${}\not={}$B;
and $p=4$: \mbox{control${}\not={}$A}, \mbox{control${}\not={}$B}. If experimental units 1
and 2 received the control treatment,~3 and 4 received treatment A, and~5
and~6 received treatment B; then \mbox{$G_{1}(i)=1$} for $i=1, \ldots, 6$;
$G_{2}(i)=1$ for $i=1,2,5,6$ and 2 for $i=3,4$; \mbox{$G_{3}(i)=1$} for
$i=1,2,3,4$ and 2 for $i=5,6$; $G_{4}(i)=i/2$ rounded up to the
nearest integer for all $i$. We will use $P$ to denote the number of
expression patterns of interest and $n_p$ to denote the number of
groups under expression pattern $p$. In the example above, $P=4$,
$n_1=1$, $n_2=n_3=2$, and $n_4=3$.

In each model, the marginal density for $\mathbf{y}_j=(y_{j1}, y_{j2},
\ldots, y_{jI})'$, the vector of observations from the $j$th gene for
$I$ experimental units, is given by $f(\mathbf{y}_j|\bt,\bolds
{\pi
})=\sum_{p=1}^P\pi_{p} f_p(\mathbf{y}_j|\bt)$, where $\bolds
{\pi
}=(\pi_{1}, \pi_{2}, \ldots, \pi_{P})'$, $\pi_{p}$ is the probability
that a gene follows expression pattern $p$, $\bt$ is a vector of
hyperparameters for the given model, and $f_p(\mathbf{y}_j|\bt)$ is the
density of $\mathbf{y}_j$ under pattern $p$ according to the given
model. The marginal likelihood of the entire data set is given by
$\prod
_{j=1}^J f(\mathbf{y}_j|\bt,\bolds{\pi})$, since observations
between genes are considered independent under each of the discussed
models. The posterior probability gene $j$ follows expression pattern
$p$ given $\mathbf{y}_j$ is $\frac{\pi_{p}f_p(\mathbf{y}_j|\bt
)}{\sum
_{p=1}^P\pi_{p}f_p(\mathbf{y}_j|\bt)}$.

For each model, estimates of $\bolds{\pi}$ and $\bt$ that maximize
the marginal likelihood can be obtained using the EM algorithm,
treating expression pattern as the unknown variable. When used,
gene-specific error variances are estimated and treated as fixed before
using the EM algorithm to estimate other model parameters. Marginal
densities and posterior probabilities are estimated by treating
parameter estimates as the true parameter values in the formulas above.

In the following subsections, we formally define four models and seven
methods of analysis. The distinguishing features of the seven methods
are summarized in Table~\ref{tab1} for future reference.

\begin{table}
\caption{Legend for method and model acronyms}\label{tab1}
\begin{tabular*}{\textwidth}{@{\extracolsep{\fill}}lcccc@{}}
\hline
&&\textbf{Relies on distinct}&&\\
&&\textbf{modeling strategies for}&\textbf{Uses}&\textbf{Accounts for}\\
&&\textbf{differences across genes}&\textbf{gene-specific}&\textbf{uncertainty in}\\
&&\textbf{and differences across}&\textbf{error variance}&\textbf{error variance}\\
\textbf{Method}&\textbf{Model}&\textbf{DE treatments}&\textbf{estimates}&\textbf{estimators}\\
\hline
LNN&LNN&&&\\
LNNMV&LNNMV&&\checkmark&\\
LNNMV*&LNNMV&&\checkmark&\\
LNNGV&LNNMV&&\checkmark&\checkmark\\
LN$^3$&LN$^3$&\checkmark&&\\
LN$^3$MV*&LN$^3$MV&\checkmark&\checkmark&\\
LN$^3$GV&LN$^3$MV&\checkmark&\checkmark&\checkmark\\
\hline
\end{tabular*}
\tabnotetext[]{}{The methods with acronyms ending in MV* use point
estimates of error variances that
account for the degrees of freedom used when
estimating treatment means (see Section~\ref{sec2.3}).}
\end{table}
%

\subsection{The lognormal--normal model}
The LNN model for the log scale observation for the $j$th gene from the
$i$th experimental unit under expression pattern $p$ can be written as
\[
y_{ji}=\mu+\tau_{jG_{p}(i)}+\varepsilon_{ji},\vadjust{\goodbreak}
\]
where
\[
\tau_{j1},
\ldots, \tau_{jn_p} \stackrel{\mathrm{i.i.d.}}{\sim} N(0,\sigma
_{\tau}^2)\quad\mbox{and}\quad \varepsilon_{j1}, \ldots, \varepsilon_{jI} \stackrel
{\mathrm{i.i.d.}}{\sim} N(0,\sigma^2).
\]
In this expression, $\mu$ represents the average expression of all
genes and groups, $\tau_{jG_p(i)}$ represents a random group effect for
observations from the $G_{p}(i)$th group (under pattern $p$) in the
$j$th gene, and $\varepsilon_{ji}$ represents a random error.

Under this model, $f_p(\mathbf{y}_j|\bt)$ is the density from a
multivariate normal distribution with mean vector $(\mu, \ldots, \mu)'$
and pattern specific covariance matrix $\Sigma_{p}=\sigma^2\I+\sigma
_{\tau}^2M_{p}$, where $\I$ is the identity matrix and $M_{p}$ is a
symmetric matrix with element $[i,j]=1$ if experimental units $i$ and
$j$ are in the same group under pattern $p$ and $[i,j]=0$ if
experimental units $i$ and $j$ are in different groups. This model has
hyperparameters $\bt=(\mu, \sigma^2, \sigma_{\tau}^2)$.

\subsection{The lognormal--normal--normal model}
To explicitly model gene effects separately from treatment effects, we
propose a three-level hierarchical model, which we denote LN$^3$. Under
the LN$^3$ model, the log scale observation from the $j$th gene and the
$i$th experimental unit under expression pattern $p$ is modeled
as
\[
y_{ji}=\mu+\gamma_j+\tau_{jG_{p}(i)}+\varepsilon_{ji},
\]
where
\begin{eqnarray*}
&\displaystyle\gamma_j \stackrel{\mathrm{i.i.d.}}{\sim} N(0,\sigma_{\gamma}^2),
\qquad\tau_{j1}, \ldots, \tau_{jn_p} \stackrel{\mathrm{i.i.d.}}{\sim}
N(0,\sigma_{\tau}^2)\quad \mbox{and}&\\
&\varepsilon_{j1}, \ldots,
\varepsilon
_{jI} \stackrel{\mathrm{i.i.d.}}{\sim} N(0,\sigma^2).&
\end{eqnarray*}
In this expression, $\mu$ represents the average expression of all
genes and groups, $\gamma_j$ represents a random gene effect for the
$j$th gene, $\tau_{jG_p(i)}$ represents a random group effect for
observations from the $G_{p}(i)$th group (under pattern $p$) in the
$j$th gene, and $\varepsilon_{ji}$ represents a random error.
Under expression pattern $p$, the density for the vector of log-scale
observations for the $j$th gene, $f_p(\mathbf{y}_j|\bt)$, is evaluated
according to a multivariate normal distribution with mean vector $(\mu,
\ldots, \mu)'$ and pattern specific covariance matrix $\Sigma
_{p}=\sigma
^2\I+\sigma_{\gamma}^2\J+\sigma_{\tau}^2M_{p}$, where $\I$ is the
identity matrix, $\J$ is a matrix of 1's, and
\[
M_{p}[i,j]=\cases{
1, &\quad $\mbox{if $G_{p}(i)=G_{p}(j)$,}$ \vspace*{2pt}\cr
0, & \quad$\mbox{otherwise}.$}
\]
This model has hyperparameters $\bt=(\mu, \sigma^2, \sigma
_{\tau}^2, \sigma_{\gamma}^2)$ and is a generalization of the LNN
model. That is, the LNN model is a special case of the LN$^3$ model in
which $\sigma_{\gamma}^2=0$.

\subsection{The lognormal--normal model with gene-specific error variances}\label{sec2.3}
The LNN model assumes that all genes have a common error variance,
$\sigma^2$. This assumption can be relaxed to allow each gene to have a
unique error variance, $\sigma_j^2$, forming the LNNMV model. We
consider three methods based on this model, including EBarrays' LNNMV.

Under this model, the log scale observation for the $j$th gene from the
$i$th experimental unit under expression pattern $p$ can be written as
\[
 y_{ji}=\mu+\tau_{jG_{p}(i)}+\varepsilon_{ji},
\]
where
\[
\tau_{j1},
\ldots, \tau_{jn_p} \stackrel{\mathrm{i.i.d.}}{\sim} N(0,\sigma
_{\tau}^2) \quad \mbox{and}\quad \varepsilon_{j1}, \ldots, \varepsilon_{jI} \stackrel
{\mathrm{i.i.d.}}{\sim
} N(0,\sigma_j^2).
\]
This model has hyperparameters $\bt=(\mu, \sigma_{\tau}^2,
\bolds
{\sigma}^2)$, where $\bolds{\sigma}^2=(\sigma_1^2,\sigma
_2^2,\ldots
,\sigma_J^2)$.

The LNNMV method from EBarrays places a scaled inverse chi-squared
distribution on the gene-specific error variances. That is, $\sigma_j^2
\sim$ inv-$\chi^2$ ($\mathrm{df}=\nu$, $\mathrm{scaling}=\Phi$), such that $\nu\Phi
/\sigma_j^2 \sim\chi^2_{\nu}$.\vspace*{-3pt} Given estimates $\hat{\nu}$ and
$\hat
{\Phi}$, the gene-specific error variances are estimated by\vspace*{1pt} $\hat
{\sigma
}_j^2=\frac{\hat{\nu}\hat{\Phi}+(I-T)S_j^2}{\hat{\nu}+I-2}$, where
$S_j^2$ is the ordinary sample variance estimator with $(I-T)$ degrees
of freedom for the log-scale observations from the $j$th gene and $T$
is total number of experimental conditions.

The denominator of the LNNMV point estimator for $\sigma_j^2$ does not
account for degrees of freedom used when estimating treatment means for
each gene in the computation of $S_j^2$. Similar to MLEs for $\sigma^2$
in a traditional ANOVA analysis, this estimator systematically
underestimates $\sigma_j^2$, resulting in liberal detection of
differential expression. If one were to use a point estimator for
$\sigma_j^2$,\vadjust{\goodbreak} we would recommend the less liberal approach of using the
posterior expectation
$\hat{\sigma}_j^2=\hat{E}(\sigma_j^2|S_j^2)=\frac{\hat{\nu}\hat
{\Phi
}+(I-T)S_j^2}{\hat{\nu}+(I-T)-2}$. We denote this approach as LNNMV*;
however, this adjusted denominator does not provide a fully adequate solution.

The EBarrays methods estimate the posterior probability that gene $j$
follows expression pattern $p$ given $\mathbf{y}_j$ as $\frac{\hat
{\pi
}_{p}f_p(\mathbf{y}_j|\hat{\bt})}{\sum_{p=1}^P\hat{\pi
}_{p}f_p(\mathbf
{y}_j|\hat{\bt})}$, assuming all hyperparameter estimates are the true
hyperparameter values. This expression is clearly sensitive to $\hat
{\bt
}$. Given that $\mu$ and $\sigma_\tau^2$ are assumed to be the same for
all genes and there are typically thousands of genes in a microarray
data set, the effective sample size for estimating these parameters is
high so that there will generally be little uncertainty associated with
the ML estimates $\hat{\mu}$ and $\hat{\sigma}_\tau^2$ obtained from
the EM algorithm. Therefore, it may be reasonable to act as if $\hat
{\mu
}=\mu$ and $\hat{\sigma}_\tau^2=\sigma_\tau^2$ when estimating
posterior pattern probabilities. Similarly, it may also be reasonable
to ignore uncertainty in the estimator of $\sigma^2$ under the LNN and
LN$^3$ models. However, when $\sigma_j^2$ is allowed to vary from gene
to gene, there will be nonnegligible uncertainty in the corresponding
estimators~$\hat{\sigma}_j^2$, which is not taken into account by
assuming $\hat{\sigma}_j^2=\sigma_j^2$. Under a model allowing for
gene-specific error variances, a better estimator of the posterior
probability that gene $j$ follows expression pattern $p$ is $\frac
{\hat
{\pi}_{p}f_p(\mathbf{y}_j|\hat{\mu},\hat{\sigma}_\tau^2,\hat
{\nu},\hat
{\Phi})}{\sum_{p=1}^P\hat{\pi}_{p}f_p(\mathbf{y}_j|\hat{\mu
},\hat{\sigma
}_\tau^2,\hat{\nu},\hat{\Phi})}$, where $f_p(\mathbf{y}_j|\hat
{\mu},\hat
{\sigma}_\tau^2,\hat{\nu},\hat{\Phi})=\int f_p(\mathbf{y}_j|\hat
{\mu
},\hat{\sigma}_\tau^2,\sigma_j^2)f(\sigma_j^2|\hat{\nu},\hat
{\Phi})
\,d\sigma_j^2$, where $f(\sigma_j^2|\hat{\nu},\hat{\Phi})$ is the
empirically estimated inverse chi-squared prior distribution for
$\sigma_j^2$.

Our suggested approach is to estimate $\hat{\nu}$ and $\hat{\Phi}$
using the method described by \citet{Smyth} and compute shrunken
estimates $\hat{\sigma}_j^2=\hat{E}(\sigma_j^2|S_j^2)$ to use when
fitting the EM algorithm to obtain estimates for $\mu, \sigma_\tau^2,
\mbox{ and } \bolds{\pi}$. Then when estimating the posterior
expression pattern probabilities for each gene, we suggest empirically
approximating $f_p(\mathbf{y}_j|\hat{\mu},\hat{\sigma}_\tau
^2,\hat{\nu
},\hat{\Phi})$ as $\sum_{q=1}^Q f_p(\mathbf{y}_j|\hat{\mu},\hat
{\sigma
}_\tau^2,\break\sigma*_q^2)/Q$ where $\sigma*_q^2$ is the $q/(Q+1)$th
quantile of $f(\sigma_j^2|\hat{\nu},\hat{\Phi})$ and $Q$ is
a~reasonably large number like 1000. We denote this method as LNNGV,
which has hyperparameters $\bt=(\mu, \sigma_{\tau}^2, \nu, \Phi
)$. The
effectiveness and impact of this suggestion are examined in Sections~\ref{s:Data Analysis}
and~\ref{s:Simulation Study}.

\subsection{The lognormal--normal--normal model with gene-specific error
variances}
As with the LNN model, the LN$^3$ model assumes that all genes have a
common error variance, $\sigma^2$, and this assumption can be relaxed
to form the LN$^3$MV model, which allows for gene-specific error
variances. For the LN$^3$MV model, we consider two methods, denoted
LN$^3$MV* and LN$^3$GV, which incorporate gene-specific error variances
in exactly the same way as the LNNMV* and LNNGV methods, respectively.
The LN$^3$MV* (LN$^3$GV) method is a generalization of the LNNMV*
(LNNGV) method. That is, the LNNMV* (LNNGV) method is a special case of
the LN$^3$MV* (LN$^3$GV) method in which $\sigma_{\gamma}^2=0$.

\section{Evidence supporting need for three-level hierarchical models}
\label{s:Evidence Supporting Need for Three-level Hierarchical Models}

Observations from a common gene are correlated for many reasons, even
across differentially expressed treatments. Variability from gene to
gene in several factors contributes to such correlation, including
binding affinity of probe sets [\citet{Binder}], amount of florescent
dye that binds to each cDNA fragment [\citet{Binder}], RNA
transcription and degradation rate [\citet{Selinger}], and the
function of genes' corresponding proteins. These considerations imply
that models for microarray data should contain gene effects like those
present in the LN$^3$ and LN$^3$MV models but omitted from the models
of Kendziorski et~al. (\citeyear{Kendziorski}).

The theoretical impact of gene effects when detecting DE genes can be
demonstrated by comparing the modeled variance of differences between
pairs of observations in two scenarios. The first scenario is when the
observations in a pair come from different groups in a common gene. The
second scenario is when the observations in a pair come from different
genes. Under the LNN model, the variance of the difference for both
scenarios is $2(\sigma_{\tau}^2+\sigma^2)$. That is, the LNN model
expects differences among same-gene observations from differentially
expressed groups to ``look like'' differences among observations from
different genes. However, when a gene effect is present, the variance
for differences between observations from different genes is $2(\sigma
_{\gamma}^2+\sigma_{\tau}^2+\sigma^2)$, which is greater than the
variance for differences between observations from different groups in
a common gene, $2(\sigma_{\tau}^2+\sigma^2)$. In this case, the LNN
model expects within-gene differences due to differential expression to
be more extreme than they actually are, which reduces the model's power
to detect differential expression. Creating a three-level hierarchical
model by adding normally distributed gene effects is a tractable and
effective method to correct this shortcoming. A similar argument can be
made when considering models that accommodate gene-specific error variances.

If information about DE groups for each gene were known for real
microarray data, we could check for evidence of gene effects by
comparing the variance of between-gene differences to the variance of
within-gene differences across DE groups. Because information about DE
groups is unknown, such a simple strategy is not possible. However, we
can fit three-level models to actual microarray data and examine the
resulting estimates of $\sigma^2_\gamma$. Because the two-level models
are special cases of three-level models with $\sigma^2_\gamma=0$,
estimates of $\sigma^2_\gamma$ far from 0 provide evidence in favor of
our proposed three-level hierarchy over the two-level hierarchy. The
next section presents results of two example microarray experiments
where the estimates of $\sigma^2_\gamma$ provide clear support for the
three-level hierarchy. We describe this point in detail in the
supplementary material [\citet{Lund}].

As additional evidence of the inadequacy of models that omit gene
effects, we compare the correlation structure implied by the LNN model
to the correlation structure present in actual microarray data.\vadjust{\goodbreak}

Under the LNN model,
\[
\operatorname{cov}(y_{ji},y_{ji'})=\cases{
\sigma_{\tau}^2, &\quad $\mbox{if $y_{ji}$ and $y_{ji'}$ are EE,}$\vspace*{2pt}\cr
0, &\quad $\mbox{otherwise}.$}
\]
For any two experimental units, under the LNN model, $\sum
_{j=1}^J \operatorname{cov}(y_{ji},y_{ji'})/\break J= \pi_{\mathrm{EE}}(i,i')\sigma_{\tau}^2$, where
$\pi_{\mathrm{EE}}(i,i')$ is the proportion of genes that are EE between
experimental units $i$ and $i'$. If experimental units $i$ and $i'$
correspond to the same experimental condition, an unbiased estimator of
$\sigma_{\tau}^2$ is given by $\hat{\sigma}_{\tau}^2(i,i')=\sum_{j=1}^J
(y_{ji}-\bar{y}_{\cdot i})(y_{ji'}-\bar{y}_{\cdot i'})/(J-1)$, because
$\pi_{\mathrm{EE}}(i,i')=1$ in this case. It follows that $\bar{\sigma}_{\tau
}^2$ is also an unbiased estimator of $\sigma^2_\tau$, where $\bar
{\sigma}_{\tau}^2$ is the average of $\hat{\sigma}_{\tau}^2(i,i')$ over
all pairs of experimental units $(i,i')$ such that the experimental
condition associated with experimental units $i$ and $i'$ is the same.

In practice, given an estimate $\hat{\pi}_{\mathrm{EE}}(i,i')$, observed
covariances between experimental units associated with different
experimental conditions are often much larger than $\hat{\pi
}_{\mathrm{EE}}(i,i')\bar{\sigma}_{\tau}^2$. Table~\ref{tab2} summarizes this phenomenon
for various treatment comparisons within two separate microarray data
sets, which are described in Section~\ref{s:Data Analysis}. Each data set was analyzed with
the LIMMA package for R developed by \citet{Smyth}. Estimates of $\pi
_{\mathrm{EE}}(i,i')$ were obtained by applying the method of \citet{Nettleton}
to the distribution of $p$-values for each pairwise comparison.
The final column provides estimates of between-treatment covariances,
which were computed as the average of all the pairwise covariances
involving one experimental unit from each of the two treatments. The
LNN and LNNMV models imply the observed between-treatment covariances
should closely match $\hat{\pi}_{\mathrm{EE}}\bar{\sigma}_{\tau}^2$, but
Table~\ref{tab2}
shows that the estimated between-treatment covariances were larger than
$\hat{\pi}_{\mathrm{EE}}\bar{\sigma}_{\tau}^2$ for every treatment comparison.

%
\begin{table}
\caption{Empirical evidence for presence of gene effects}\label{tab2}
\begin{tabular*}{\textwidth}{@{\extracolsep{\fill}}lcccc@{}}
\hline
\textbf{Dataset (conditions)}& $\bolds{\hat{\pi}_{\mathrm{EE}}}$& $\bolds{\bar{\sigma}_{\tau}^2}$
&$\bolds{\hat{\pi}_{\mathrm{EE}}\bar{\sigma}_{\tau}^2}$ &\textbf{Average across condition cov} \\
\hline
DC3000 (NaCl, ctrl)&0.716&0.952&0.681&0.903\\
DC3000 (phen, ctrl)&0.693&0.977&0.677&0.910\\
DC3000 (PEG, ctrl)&0.352&0.914&0.322&0.838\\
DC3000 (H$_2$O$_2$, ctrl)&0.961&0.957&0.920&0.948\\
Mouse (Ch, FF)&0.874&0.281&0.245&0.280\\
Mouse (Ch, MP)&0.824&0.281&0.231&0.279\\
Mouse (FF, MP)&0.956&0.284&0.272&0.284\\
\hline
\end{tabular*}
\end{table}

The additional covariance observed between experimental units from
different experimental conditions is easily explained by the presence
of gene effects. For any two experimental units, under the LN$^3$
model, $\sum_{j=1}^J \operatorname{cov}(y_{ji},\break y_{ji'})/ J=\sigma_{\gamma}^2+\pi
_{\mathrm{EE}}(i,i')\sigma_{\tau}^2$ rather than $\pi_{\mathrm{EE}}(i,i')\sigma_{\tau}^2$.

\section{Data analysis}
\label{s:Data Analysis}

\subsection{Data set descriptions}

We analyzed a NimbleGen mRNA data set of 5608 genes from the DC3000
strain of the bacterial plant pathogen \textit{Pseudomonas syringae},
resulting from an unpublished experiment conducted in the Department of
Plant Pathology at Iowa State University. NimbleGen performed RMA
normalization on the data [\citet{Irizarry}]. The experiment had two
biological replicate samples grown in each of five different media:
control (ctrl), phenol (phen), sodium chloride (NaCl), polyethylene
glycol MW8000 (PEG), and hydrogen peroxide (H$_2$O$_2$). Before
analyzing the data, the primary investigator suggested that any two
noncontrol media will be EE only when they are also EE with the
control, which reduces the number of expression patterns included in
the analysis. Because each of the four treatments can be either EE or
DE with the control, there are $2^4=16$ different expression patterns
to consider.

The second data set we analyzed is a subset of the data used in \citet{Somel},
available at the Gene Expression Omnibus (GEO) website as
GDS3221. This experiment examined the impact of diet on the expression
of 45,101 genes in mice. We analyzed data from nine Affymetrix GeneChips
corresponding to three treatment groups of three mice each. Each
treatment involved ad libidum feeding of one of the following diets:
(1) vegetables, fruit, and yogurt identical to the diet fed to
chimpanzees in their ape facility (Ch);
(2) McDonald's fast food (FF);
(3)~mouse pellets on which the mice were raised (MP). To keep the
presentation simple, we have omitted data from a second batch of chips
and a fourth diet group (cafeteria food), which produced expression
profiles very similar to those from the McDonald's diet. With the three
included treatment groups, there are a total of five possible
expression patterns: Ch${}={}$FF${}={}$MP; Ch${}={}$FF${}\neq{}$MP; Ch${}\neq{}$FF${}={}$MP;
Ch${}={}$MP${}\neq{}$FF; and Ch${}\neq{}$FF, Ch${}\neq{}$MP, FF${}\neq{}$MP.

%
\begin{table}
\caption{Hyperparameter estimates and estimated proportion of null genes
for DC3000 (top) and~mouse diet (bottom) data from each of the models}\label{tab3}
\begin{tabular*}{\textwidth}{@{\extracolsep{\fill}}ld{3.4}d{1.5}d{1.5}d{1.5}d{1.5}@{}}
\hline
&\multicolumn{5}{c@{}}{\textbf{Model used to analyze}}\\[-6pt]
&\multicolumn{5}{c@{}}{\hrulefill}\\
\textbf{Parameter}&\multicolumn{1}{c}{\textbf{GG}}&\multicolumn{1}{c}{\textbf{LNN}}&\multicolumn{1}{c}{\textbf{LNNGV}}&
\multicolumn{1}{c}{\textbf{LN}$^{\bolds{3}}$}&\multicolumn{1}{c@{}}{\textbf{LN}$^{\bolds{3}}$\textbf{GV}}\\
\hline
$\hat{\alpha}$&69.8&\multicolumn{1}{c}{--}&\multicolumn{1}{c}{--}&\multicolumn{1}{c}{--}&\multicolumn{1}{c}{--}\\
$\hat{\alpha}_0$&1.54&\multicolumn{1}{c}{--}&\multicolumn{1}{c}{--}&\multicolumn{1}{c}{--}&\multicolumn{1}{c}{--}\\
$\hat{\nu}$*&0.0254&\multicolumn{1}{c}{--}&\multicolumn{1}{c}{--}&\multicolumn{1}{c}{--}&\multicolumn{1}{c}{--}\\
$\hat{\mu}$&\multicolumn{1}{c}{--}&0.501&0.419&0.277&0.264\\
$\hat{\sigma}_{\tau}^2$&\multicolumn{1}{c}{--}&0.982&0.878&0.151&0.101\\
$\hat{\sigma}_{\gamma}^2$&\multicolumn{1}{c}{--}&\multicolumn{1}{c}{--}&\multicolumn{1}{c}{--}&0.813&0.832\\
$\hat{\sigma}^2$&\multicolumn{1}{c}{--}&0.0129&\multicolumn{1}{c}{--}&0.0116&\multicolumn{1}{c}{--}\\
$\hat{\Phi}$&\multicolumn{1}{c}{--}&\multicolumn{1}{c}{--}&0.00509&\multicolumn{1}{c}{--}&0.00509\\
$\hat{\nu}$&\multicolumn{1}{c}{--}&\multicolumn{1}{c}{--}&3.546&\multicolumn{1}{c}{--}&3.546\\
$\hat{\pi}_{\mathit{null}}$&0.728&0.721&0.657&0.655&0.492\\[3pt]
$\hat{\alpha}$&269.5&\multicolumn{1}{c}{--}&\multicolumn{1}{c}{--}&\multicolumn{1}{c}{--}&\multicolumn{1}{c}{--}\\
$\hat{\alpha}_0$&4.59&\multicolumn{1}{c}{--}&\multicolumn{1}{c}{--}&\multicolumn{1}{c}{--}&\multicolumn{1}{c}{--}\\
$\hat{\nu}$*&0.0187&\multicolumn{1}{c}{--}&\multicolumn{1}{c}{--}&\multicolumn{1}{c}{--}&\multicolumn{1}{c}{--}\\
$\hat{\mu}$&\multicolumn{1}{c}{--}&0.206&0.210&0.194&0.194\\
$\hat{\sigma}_{\tau}^2$&\multicolumn{1}{c}{--}&0.279&0.281&0.00468&0.00678\\
$\hat{\sigma}_{\gamma}^2$&\multicolumn{1}{c}{--}&\multicolumn{1}{c}{--}&\multicolumn{1}{c}{--}&0.278&0.275\\
$\hat{\sigma}^2$&\multicolumn{1}{c}{--}&0.00346&\multicolumn{1}{c}{--}&0.00331&\multicolumn{1}{c}{--}\\
$\hat{\Phi}$&\multicolumn{1}{c}{--}&\multicolumn{1}{c}{--}&0.00249&\multicolumn{1}{c}{--}&0.00249\\
$\hat{\nu}$&\multicolumn{1}{c}{--}&\multicolumn{1}{c}{--}&8.186&\multicolumn{1}{c}{--}&8.186\\
$\hat{\pi}_{\mathit{null}}$&0.958&0.954&0.931&0.802&0.840\\
\hline
\end{tabular*}
\end{table}

\subsection{Analysis of real data}

We analyzed these data sets with each of the eight methods and report
the resulting parameter estimates from the GG, LNN, LNNGV, LN$^3$, and
LN$^3$GV methods in Table~\ref{tab3}. [The LNNMV*, LNNMV, and LN$^3$MV* methods
share theoretical models (and thus parameter estimates) with the LNNGV,
LNNGV, and LN$^3$GV methods, resp.] The parameter estimates in
Table~\ref{tab3} are consistent with what we expected. For both data sets, when
a random gene effect is accounted for in the model, the estimated
treatment effect variance decreases drastically and the gene effect
variance is estimated to be much larger than the treatment effect
variance. This means the LN$^3$ and LN$^3$GV methods are able to detect
smaller treatment effects than their respective two-level counterparts,
LNN and LNNGV. It is not surprising then to see that for both data sets
the LNN method estimates a larger proportion of genes following the
null pattern than does the LN$^3$ method, or that the LNNGV method
estimates a larger proportion of genes following the null pattern than
does the LN$^3$GV method.

\begin{figure}

\includegraphics{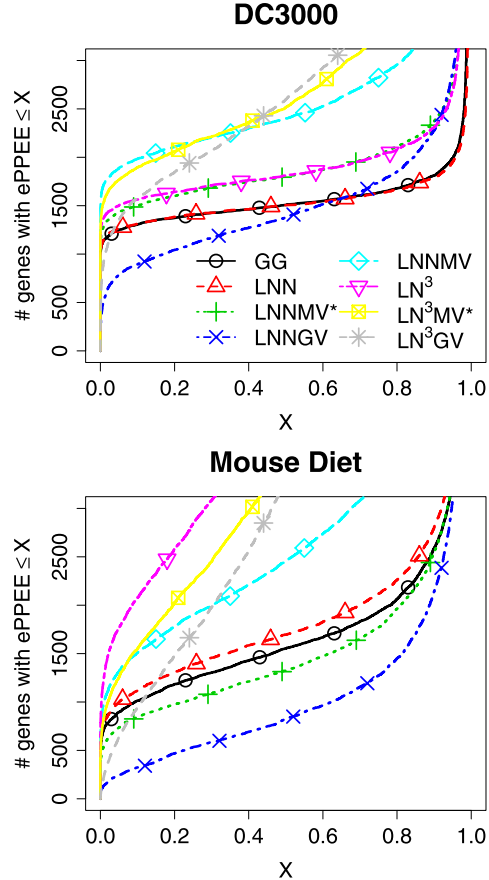}

  \caption{Comparison across methods of empirical ePPEE CDFs for DC3000
(top) and mouse diet (bottom) data.}\label{fig1}
\end{figure}

Rather than examining parameter estimates, researchers are often more
interested in creating lists of genes that are likely to follow
expression patterns of interest. To construct a list of DE genes, one
would collect all genes with an estimated posterior probability of
equivalent expression (ePPEE) less than a given threshold. When the
ePPEE falls below the given threshold for many genes, not all
identified potentially DE genes may be individually studied further.
However, the size and contents of the entire list provide important
information to researchers about the global effects of the treatments
on gene expression. The composition of a long list of potentially DE
genes forms the basis for popular gene set enrichment analyses that are
commonly used to interpret the results of microarray experiments. To
examine the practical differences between gene lists created by the
methods, we begin by plotting the empirical CDF of the ePPEEs for each
method for the two data sets in Figure~\ref{fig1}. These plots quickly provide
the observed size of a gene list for any PPEE cutoff, obtained by
intersecting a vertical line at the desired PPEE cutoff with the curve
for each method.

The plots show substantial differences between the examined methods in
the number of detected genes over a wide range of PPEE thresholds. For
models with gene-specific error variances, incorporating uncertainty in
estimated error variances greatly reduced the number of detected genes
(LNNGV and LN$^3$GV curves are
lower than LNNMV* and LN$^3$MV* curves, resp.). In the DC3000 data at a PPEE cutoff of
0.1, for example, the LNNMV, LNNMV*, and LNNGV methods would produce
lists with 1983, 1498, and 893 genes, respectively. Incorporating gene
effects greatly increased the number of detected genes (LN$^3$, LN$^3$MV*, and
LN$^3$GV curves are higher than LNN, LNNMV*, and LNNGV
curves, resp.). In the mouse
diet data at a PPEE cutoff of 0.1, for example, the LN$^3$GV method
identified almost three times as many DE genes as the LNNGV method (945
vs. 324 genes, resp.).
These results indicate that differences between the methods' ePPEEs are
practically significant, and care should be taken when choosing among
the suggested methods.

Constraints on time, money, material, and personnel resources limit the
number of genes that researchers will follow up on with further study.
Thus, the overlap between lists from each method containing a fixed
number of the most significant genes is an important feature for
assessing the similarity between methods' results. Table~\ref{tab4} provides the
size of pairwise intersections of lists containing the 200 most
significant genes from each method for the DC3000 and mouse diet data
sets, respectively. These results show substantial practical
differences between rankings, as many lists overlap by roughly half
their genes and most lists overlap by fewer than 150 genes.

%
\begin{table}
\caption{Overlap in lists of top 200 most significant DE genes for
DC3000 (top)~and~mouse diet (bottom) data}\label{tab4}
\begin{tabular*}{\textwidth}{@{\extracolsep{\fill}}lccccccc@{}}
\hline
\textbf{Method}&\textbf{1}&\textbf{2}&\textbf{3}&\textbf{4}&\textbf{5}&\textbf{6}&\textbf{7}\\
\hline
(1) GG&200&&&&&&\\
(2) LNN&187&200&&&&&\\
(3) LNNMV&122&119&200&&&&\\
(4) LNNMV*&118&120&160&200&&&\\
(5) LNNGV&130&127&185&162&200&&\\
(6) LN$^3$&186&198&117&118&125&200&\\
(7) LN$^3$MV*&117&114&194&154&184&113&200\\
(8) LN$^3$GV&\phantom{0}77&\phantom{0}81&137&149&133&\phantom{0}79&135\\[3pt]
(1) GG&200&&&&&&\\
(2) LNN&193&200&&&&&\\
(3) LNNMV&108&107&200&&&&\\
(4) LNNMV*&125&124&152&200&&&\\
(5) LNNGV&\phantom{0}88&\phantom{0}87&173&136&200&&\\
(6) LN$^3$&193&197&109&124&\phantom{0}89&200&\\
(7) LN$^3$MV*&\phantom{0}93&\phantom{0}92&181&134&184&\phantom{0}94&200\\
(8) LN$^3$GV&\phantom{0}83&\phantom{0}82&155&148&158&\phantom{0}82&148\\
\hline
\end{tabular*}
\end{table}

\section{Simulation study}
\label{s:Simulation Study}

Here we briefly summarize our simulation study and its results.
Detailed accounts of simulation procedures and results are presented in
the supplementary material [\citet{Lund}].

We conduct a variety of simulations to assess the accuracy and power of
the considered methods. By ``accuracy,'' we refer to the property that
for any given collection of genes the average estimated posterior
probability for each pattern should closely match the proportion of
genes in the collection that follow the given pattern. By ``power,'' we
refer to a method's ability to detect differential expression. We
prefer the method that creates the largest list of genes for a given
ePPEE threshold, provided that the method's ePPEEs are accurate.

We simulated data from each of the five models (GG, LNN, LNNMV, LN$^3$,
and LN$^3$MV) using the model parameters reported for the DC3000 data
set in Table~\ref{tab3}. In addition to these model-based simulations, we also
conducted simulations using an HIV mRNA expression data set from the
GEO website, named GDS1449. We analyzed each simulated data set with
each method and recorded estimated posterior probabilities for each
expression pattern for each gene.

The simulation results clearly support our claims that failing to
distinctly model gene and gene-specific treatment effects reduces power
and produces conservative results and that using point estimates of
error variances produces liberal results. The LN$^3$GV method stands
out as the best method from these simulations. The LN$^3$GV method was
the only method to produce accurate ePPEEs in all simulation scenarios,
and no method produced better average significance rankings (as seen in
ROC curves) than those of the LN$^3$GV method in any simulation
scenario. The LN$^3$GV method exhibited greater power than the LNNGV
method, which was the only other method that did not produce liberal
results in at least one simulation scenario.

\section{Discussion}
\label{s:Discussion}

When modeling a data set that includes multiple observations from each
of multiple genes, a conventional analysis would begin with a model
that incorporates gene effects. One might decide to omit gene effects
if, after looking, there was no evidence of gene effects or if results
from an analysis were not affected by the omission of gene effects. We
have demonstrated that gene effects are present in real data sets and
provided generalizations of the methods based on lognormal two-level
hierarchical models to include gene effects. These generalizations
behave nearly identically to their two-level counterparts when
analyzing data without gene effects and improve power and accuracy when
data contain gene effects. These extensions serve as an example of how
other hierarchical models that omit gene effects might be improved by
more versatile modeling.

Using point estimates of gene-specific error variances without
accounting for their uncertainty produces liberal ePPEEs. We have
suggested a~corrected approach that involves integration over an
empirically estimated prior distribution for the error variances and
demonstrated this adaptation yields accurate ePPEEs.

As noted in the \hyperref[s:intro]{Introduction}, we have identified nearly a dozen methods
that omit gene effects. There are far more methods in the microarray
data analysis literature that do not suffer from this problem. Most
published methods explicitly or implicitly include gene effects whose
distribution is allowed to differ from the distribution of within gene
treatment effects. Methods based on gene-specific linear models that
make no attempt to borrow information across genes fall into this
category, as do methods that borrow\vadjust{\goodbreak} information across genes only for
the purpose of improved error variance estimation. While we expect our
LN$^3$GV method to perform well when compared against the large
collection of competing approaches, a broad comparison of methods is
beyond the scope of this paper, and we make no claims of superiority
here. Our main point is that the hierarchical modeling approach
pioneered by Kendziorski et~al. (\citeyear{Kendziorski}) can be improved by the inclusion of
both gene and gene-specific treatment effects. Given the influential
nature of the original work of Kendziorski et~al. (\citeyear{Kendziorski}), we think this is
an important point to make.

The development of the LNN and GG models by Kendziorski et~al. (\citeyear{Kendziorski})
represents groundbreaking work on the hierarchical modeling of gene
expression data. We have shown how to improve on the original work by
allowing for random gene effects and replacing gene-specific error
variance point estimates without dramatically affecting computational
costs. Adding random gene effects to a model increases the dimension of
the parameter space across which the EM algorithm must optimize by one,
but does not substantially increase computational costs. For any of the
described methods, analyzing a data set with 5000 genes, 9 experimental
units, and 4 expression patterns of interest takes less than 10 minutes
using a single Linux machine and R code that calls a C routine to
evaluate multivariate normal densities. We have developed the R package
LN3GV (available at the CRAN webpage) to implement the LNNMV*, LNNGV,
LN$^3$, LN$^3$MV*, and LN$^3$GV methods discussed in this article.
Throughout this paper, the GG, LNN, and LNNMV methods were implemented
via the EBarrays package.

We have focused on the approach of Kendziorski et~al. (\citeyear{Kendziorski}) not only
because of its influential nature but also because of its unique and
elegant approach to inference for experiments with more than two
treatments. The vast majority of competing approaches have been
developed primarily for the case of two treatments. While it is easy to
extend many of these methods to cover the case of more than two
treatments, very few methods outside the Kendziorski et~al. (\citeyear{Kendziorski}) lineage
provide an inherent natural strategy for classifying genes according to
their pattern of expression across multiple treatments. Thus, we
believe our efforts to improve the original work of Kendziorski et~al. (\citeyear{Kendziorski}) have been well spent.



\begin{supplement}
\stitle{Additional evidence supporting need for three-level hierarchy
and simulation study details}
\slink[doi]{10.1214/12-AOAS535SUPP} 
\slink[url]{http://lib.stat.cmu.edu/aoas/535/supplement.pdf}
\sdatatype{.pdf}
\sdescription{The correlation across genes present in real
microarray data makes directly testing the statistical significance of
gene effect variance estimates intractable. We present a simulation
study that demonstrates the gene effect variance estimates obtained
when analyzing the DC3000 and mouse diet data sets are drastically
greater than those that arise when analyzing data simulated without
gene effects. We also\vadjust{\goodbreak} provide detailed accounts of simulation
procedures and results used to evaluate the considered methods. These
simulations clearly support our claims regarding the importance of
distinctly modeling gene and gene-specific treatment effects and
accounting for uncertainty in error variance estimators.\vspace*{-3pt}}
\end{supplement}


%

%

\printaddresses

\end{document}